\documentclass[a4paper, 10pt]{report}

\usepackage[cp1250]{inputenc}
\usepackage[T1]{fontenc}
\usepackage{amsfonts}
\usepackage{graphicx}
\usepackage{amsmath}

\usepackage{amssymb}

\usepackage{bibtopic}
\usepackage{color}
\usepackage{ulem}

\AtBeginDocument{}
\usepackage{etoolbox}
\patchcmd{\thebibliography}{\chapter*}{\section*}{}{}

\makeatletter
\renewcommand{\thesection}{%
  \ifnum\c@chapter<1 \@arabic\c@section
  \else \thechapter.\@arabic\c@section
  \fi
}
\makeatother

\begin{document}

{\LARGE \bf

\centerline{Effect of radiation-like solid on CMB anisotropies}

}

\vskip 1cm
\begin{center}
{Peter Mészáros}

\vskip 2mm {\it Department of Theoretical Physics, Comenius
University, Bratislava, Slovakia}

\vskip 4mm September 30, 2016 \vskip 1cm {\bf Abstract}
\end{center}

\vskip 5mm \noindent We compute the CMB angular power spectrum
in the presence of the radiation-like solid with the same pressure to energy
density ratio as for the radiation but with nonzero shear modulus.
The effect can be observable not only for large-angle anisotropies
as shown by {{\it Balek and Skovran} [2015]}
but also for very small-angle ones.

\section{Introduction}

The idea of solid matter in cosmology was used as an attempt
to give an alternative explanation of the acceleration of the universe, see
{[{\it Bucher, Spergel}, 1999]}
where the dark energy was replaced by a solid with negative pressure to
energy density ratio $w$. Further development of this theory
{[{\it Battye et al.}, 1999; {\it Leite, Martins}, 2011; {\it Battye, Moss}, 2006
; {\it Battye, Moss}, 2009; {\it Battye, Moss}, 2007; {\it Battye, Pearson}, 2013
; {\it Kumar et al.}, 2013]}
included works where inflation is driven by a solid (solid inflation)
{[{\it Gruzinov}, 2004; {\it Endlich et al.}, 2013; {\it Akhshik}, 2015
; {\it Bartolo et al., Ricciardone}, 2013; {\it Sitwell, Sigurdson}, 2013]}.
An important example of materialisation of such a solid are cosmic strings and domain walls
{[{\it Battye et al.}, 1999; {\it Leite}, 2011; {\it Kumar et al.}, 2013]}.

The effect of the presence of a solid can be also obtained by any mechanism leading
to the appropriate form of nonzero nondiagonal space components of energy-momentum tensor.
One can for instance consider the Lagrangian $\mathcal{L}[\phi]$ depending
on the fields $\phi^A(x)$, which have the meaning of {\it internal coordinates} of a solid
{[{\it Endlich et al.}, 2013; {\it Akhshik}, 2015]}.
As an extension of the parametric space of the theory
one can consider also a solid with positive pressure to energy density ratio
{[{\it Balek, Skovran}, 2014]}.
An important special case is the {\it radiation-like solid} with $w=1/3$,
which does not change the evolution of the unperturbed universe if the total energy density
of radiation and this solid equals the energy density of radiation
in the standard case. Materialisation of such a solid could be a Coulomb crystal
with relativistic Fermi gas of moving particles, or a network of speculative
'spring-like' strings with energy inversly proportional to their length.

A solid matter affects the evolution of perturbations if it appeared with flat
internal geometry and nonzero shear stress acting in it.
Such solidification cannot occur in pure radiation and must be related
to another kind of particles distributed anisotropicaly before
the solidification, which could be perhaps present in the universe as a remnant of solid inflation.

However current observations are well explained by the standard theory
including radiation, baryonic matter, cold dark matter (CDM) and dark energy
(in later stages of evolution of the universe), the future observations
could possibly appear not to be fully explained by the standard theory
and some nonstandard theories, including the model presented in this work,
may become relevant.

The effect of the radiation-like solid on the long-wavelength perturbations
and coefficients of angular power spectrum of the cosmic microwave backround (CMB) anisotropies
for very low multipole moments was studied by {{\it Balek and Skovran} [2015]}.
Since the sound speed does not appear in the solution of equations
for perturbations in the long-wavelength limit (see (7.69) in {[{\it Mukhanov}, 2014]}),
one can suppose the CDM to be coupled to the baryonic matter,
if the effect is considered only for low multipole moments.

In this work we study effects of the radiation-like solid appearing shortly
after inflation on the CMB angular power spectrum
not only for low multipole moments,
so we must investigate a model where the CDM and the baryonic matter
are decoupled. In the second section we derive equations for perturbations in this case,
in the third section we summarize effects of the presence of radiation-like solid
on the CMB anisotropies and in the last section we discuss the results.
We use the signature $(+,-,-,-)$ for the metric and units in which $c=1$ and $16\pi G=1$.

\section{Equations for perturbations}
We will apply scalar perturbation theory to
flat Friedmann--Robertson--Walker--Lema\^itre (FRWL) universe filled with
radiation, baryonic matter and CDM,
and add an elastic radiation-like matter
with nonzero shear modulus to it.
The presence of the radiation-like solid with the pressure to energy density ratio $w=1/3$,
the same as for radiation, does not change the evolution of the unperturbed universe
if the total energy density of all radiation-like components is unchanged.

We will use {\it proper-time comoving gauge} {[{\it Pol\'ak, Balek}, 2008]}
in which the $(0,0)$-component of metric tensor is unperturbed 
and the shift vector $\delta\mathbf{x}$
is zero for radiation-like solid as well as for all matter coupled with it.
In this gauge the scalar part of metric takes the form
\begin{eqnarray}
ds^{(S)2}=a^2[d\eta^2+2B_{,i}d\eta dx^i-(\delta_{ij}(1-2\psi)-2E_{,ij})dx^idx^j],
\nonumber
\label{eq:1}
\end{eqnarray}
where $a$ denotes the scale parameter, $\eta$ is the conformal time and $B$, $\psi$ and $E$
are the functions describing scalar perturbations. Considering both radiation and baryonic matter
coupled to the radiation-like solid, the scalar part of energy-momentum tensor is
\begin{eqnarray}
&T_0^{\phantom{0}0(S)}=\rho+\rho_{+}^{b+r}(3\psi+\mathcal{E})+\delta\rho^d,
\qquad
T_i^{\phantom{i}0(S)}=\rho_{+}^{b+r}B_{,i}+\rho^d\vartheta_{,i},
\nonumber\\
&T_i^{\phantom{i}j(S)}=-p\delta_i^j-K(3\psi+\mathcal{E})\delta_i^j-2\mu E^T_{,ij}
+2\mathcal{N}a^{-1}E^T_{,ij\eta },
\label{eq:2}
\end{eqnarray}
where $\rho$ is the total energy density, $\rho^d$ is the energy density of the CDM,
$\rho^{b+r}$ is the energy density of all
matter except for the CDM, $\rho_{+} = \rho + p$, $p$ is pressure, $\delta\rho^d$
is the perturbation of the CDM energy density, $\vartheta_{,i}$ is the scalar part of
the velocity (longitudinal part in Helmholtz decomposition) of CDM with respect to other matter,
$K$ is the compressional modulus of
matter, $\mu$ is the shear modulus of the radiation-like solid, $\mathcal{N}$ is the shear
viscosity coefficient due to coupling of radiation and baryonic matter before recombination
responsible for {\it Silk damping}, $\mathcal{E}=\triangle E$ and $E^T_{,ij}\equiv E_{,ij}-(1/3)E_{,kk}\delta_{ij}$
is traceless part of the tensor $E_{,ij}$.

Energy-momentum tensor with nondiagonal spatial components can be obtained (also) from
Lagrangians $\mathcal{L}[\phi]$ which are invariant under internal rotations and
translations {[{\it Endlich et al.}, 2013; {\it Akhshik}, 2015]}
\begin{eqnarray}
\phi^A \to M^A_{\phantom{A}B}\phi^B, \qquad \phi^A \to \phi^A + C^A, \qquad
M^A_{\phantom{A}B} \in SO(3), \qquad C^A\in\mathbb{R}^3, \qquad A=1,2,3, \nonumber
\label{eq:dop1}
\end{eqnarray}
where the capital indices are raised and lowered by the Euclidean metric $\delta_{AB}$.
If the object described by the theory is solid (elastic) matter,
the three-component field $\phi^A$ is interpreted
as the so-called spatial {\it internal coordinates}
which move along the solid matter, so that in the perturbed FRWL universe we have
$\phi^A(x) = \delta^A_i x^i - \delta x^A(x)$.
The fields $\delta x^A(x)$ which describe the perturbed state of the solid play the role
of Goldstone boson fields in this theory. In the proper-time comoving gauge
these boson fields disappear.
An example of Lagrangian leading to the energy-momentum tensor (\ref{eq:2})
to the first order in perturbation theory,
except for the Silk damping term, is
\begin{eqnarray}
\mathcal{L} = \mathcal{L}_0 + c_1 \partial_\mu \phi^A \partial^\mu \phi_A
+(c_2 \delta_{AB}\delta_{CD} + c_3 \delta_{AC}\delta_{BD})
\partial_\mu \phi^A \partial^\mu \phi^B \partial_\nu \phi^C \partial^\nu \phi^D,
\nonumber
\label{eq:dop3}
\end{eqnarray}
where $\mathcal{L}_0=\rho+9(K-\rho_+)/8$, $c_1=(3K-\rho_+)a^2/4$,
$c_2=(K+\rho_+-2\mu/3)a^4/8$, $c_3=(\mu-\rho_+)a^4/4$.
The Silk damping term is obtained from the expression of
energy-momentum tensor for an imperfect fluid.

For the variables $\psi$, $\mathcal{E}$, $B$, $\delta\rho^d$ and $\vartheta$
we have five equations derived from the components of Einstein field equations
and the energy-momentum conservation law,
$2G_{i}^{\phantom{i}0}=T_{i}^{\phantom{i}0}$, $2G_{0}^{\phantom{0}0}=T_{0}^{\phantom{0}0}$,
$T_{i\phantom{\mu};\mu}^{\phantom{i}\mu}=0$ for all matter except for the CDM and the CDM respectively
and $T_{0\phantom{\mu};\mu}^{\phantom{0}\mu}=0$ for the CDM;
in the first order of the perturbation theory
in the case when the internal geometry of the radiation-like solid is flat.
The functions $B$, $\delta\rho^d$ and $\vartheta$ are not defined uniquely,
because the proper-time comoving gauge allows for a residual transformation
$\eta\to\eta+a^{-1}\delta t(\mathbf{x})$, where $\delta t(\mathbf{x})$ is a local shift
of the moment at which the time count starts.
The function $\mathcal{E}$ is invariant under this residual tranformation and $B$, $\psi$,
CDM energy density contrast $\delta=\delta\rho^d/\rho^d$ and $\vartheta$ can be rewriten as
\begin{eqnarray}
B = \mathcal{B}+\chi, \qquad \psi=-\mathcal{H}\chi,
\qquad \delta=\hat{\delta}-3\mathcal{H}\chi, \qquad
\vartheta=\hat{\vartheta}+\chi,
\nonumber
\label{eq:8}
\end{eqnarray}
where $\chi$ transforms as $\chi\to\chi+a^{-1}\delta t(\mathbf{x})$ and $\mathcal{B}$,
$\hat{\delta}$ and $\hat{\vartheta}$
are invariant.

It is convenient to introduce a new time $\zeta=\eta/\eta_*$,
where $\eta_*=\eta_{eq}/(\sqrt{2}-1)$ and $\eta_{eq}$ denotes
the moment when the energy density of matter is
equal to the energy density of radiation.
When the contribution of
dark energy to the total energy density is negligible, which is the case before
recombination, the scale parameter $a$ can be written in terms of $\zeta$ as $a=a_{eq}\zeta(\zeta+2)$.
The invariant combinations of equations for perturbations
of the form of a plane wave with the comoving wave vector $\mathbf{k}$ are
\begin{eqnarray}
\mathcal{E}'&=&-(s^2+3\alpha_1\tilde{\mathcal{H}}^2)\tilde{\mathcal{B}}
-\alpha_1\tilde{\mathcal{H}}\mathcal{E} - 3\alpha_2\tilde{\mathcal{H}}^2\theta
-\alpha_2\tilde{\mathcal{H}}\hat{\delta},
\label{eq:10}\\
\tilde{\mathcal{B}}'&=&(3c_{s0}^2+\alpha_1-1)\tilde{\mathcal{H}}\tilde{\mathcal{B}}+
c_{s||}^2\mathcal{E}+\alpha_2\tilde{\mathcal{H}}\theta+\sigma\mathcal{E}',
\label{eq:11}\\
\theta'&=&\alpha_1\tilde{\mathcal{H}}\tilde{\mathcal{B}}+(\alpha_2-1)\tilde{\mathcal{H}}\theta,
\label{eq:12}\\
\hat{\delta}'&=&\mathcal{E}'+s^2\tilde{\mathcal{B}}-s^2\theta,
\label{eq:13}
\end{eqnarray}
where the prime denotes differentiation with respect to $\zeta$,
$\tilde{\mathcal{B}}=\mathcal{B}/\eta_{*}$,
$\theta=\hat{\vartheta}/\eta_{*}$, $s=k\eta_{*}$, $\alpha_1=3\rho_{+}^{b+r}/(2\rho)$,
$\alpha_2=3\rho^d/(2\rho)$, $\tilde{\mathcal{H}}=a^{-1}da/d\zeta$,
$\sigma=4\mathcal{N}/(3a\rho_{+}^{b+r}\eta_{*})$, $c_{s0}$ is the auxiliary sound speed and
$c_{s||}$ the longitudinal sound speed. The two sound speeds are defined as
\begin{eqnarray}
\rho_{+}^{b+r}c_{s0}^2=K, \qquad
c_{s||}^2=c_{s0}^2+\frac{4}{3}\frac{\mu}{\rho_{+}^{b+r}}=(1+3\xi)c_{s0}^2,
\nonumber
\label{eq:m14b}
\end{eqnarray}
where $\xi$ is the dimensionless shear stress parameter defined as
\begin{eqnarray}\label{eq:m14}
\xi\equiv\frac{\mu}{\rho^r},
\nonumber
\end{eqnarray}
and $\rho^r$ denotes sum of energy densities of radiation and radiation-like solid.
Equations (\ref{eq:10})-(\ref{eq:13}) are generalization of (4) in
{[{\it Balek, Skovran}, 2015]},
valid in long-wavelength limit.

Instead of the proper-time comoving gauge, one often uses {\it Newtonian gauge}
in which the scalar part of space-time metric is diagonal.
Metric in the Newtonian gauge can be described by
two potentials invariant under coordinate transformations,
$\Phi$, called {\it Newtonian potential}, and $\Psi$,
as $g_{00}=a^2(1+2\Phi)$ and $g_{ij}=-a^2(1-2\Psi)\delta_{ij}$.
The functions $\Phi$ and $\Psi$ can be rewritten as
\begin{eqnarray}
\Psi=\mathcal{H}(\mathcal{B}-E'), \qquad
\Phi=\Psi-\mu a^2E, \label{eq:14}
\nonumber
\end{eqnarray}
where the difference between $\Phi$ and $\Psi$ is given by the traceless part
of $2G_{i}^{\phantom{i}j}=T_{i}^{\phantom{i}j}$.
The physical CDM energy density contrast $\bar{\delta}$ and function
$\bar{\vartheta}$ describing
scalar part of the
relative physical velocity of the CDM with respect to the rest of the matter
used in Newtonian gauge are
\begin{eqnarray}
\bar{\delta}=\hat{\delta}+3\mathcal{H}(\mathcal{B}-E'), \qquad
\bar{\vartheta}=\hat{\vartheta}+\mathcal{B}-E'. \label{eq:15}
\nonumber
\end{eqnarray}
Another useful relation,
\begin{eqnarray}
\overline{\delta\rho}^{b+r}=\rho_{+}^{b+r}(3\Psi+\mathcal{E}),
\label{eq:16}
\end{eqnarray}
is valid also for baryonic matter and radiation including solid separately.

\section{CMB power spectrum}
The CMB anisotropies are given by fluctuations of radiating matter density
at the time of last scattering and effects influencing photons
during their propagation to the observer.
Considering the Sachs--Wolfe effect, Doppler effect
and the finite thickness effect,
the relative fluctuations of the CMB temperature can
be written as
\begin{eqnarray}
\frac{\delta T}{T}(\eta_0,\mathbf{l})=\int \frac{d^3k}{(2\pi)^{3/2}}
\left[
\left(
\alpha(\mathbf{k})
+\beta(\mathbf{k})\frac{\partial}{\partial\eta_0}
\right)_{\eta_r} e^{i\mathbf{k}\cdot\mathbf{l}(\eta_r-\eta_0)}
\right] e^{-\Sigma k^2},
\label{eq:b10}
\end{eqnarray}
where $\mathbf{l}$ is the unit vector pointing from the place on sky from which the radiation is coming
towards the observer, $\eta_0$ and $\eta_r$ denote the conformal time today and at the time of recombination
respectively,
$\Sigma\equiv(6\kappa^2\mathcal{H}(\eta_r)^2)^{-1}$,
$\kappa$ being 
the ratio of the ionization energy of the 2S state of the hydrogen
atom and the energy $k_B T_r$ corresponding to the temperature of
recombination $T_r$ (see \S 3.6.3 in {[{\it Mukhanov}, 2014]}),
and $\alpha(\mathbf{k})$ and $\beta(\mathbf{k})$ are defined as
\begin{eqnarray}
\alpha(\mathbf{k})=\frac{1}{4}\delta_{\gamma\mathbf{k}}+\Phi_{\mathbf{k}},
\quad
\beta(\mathbf{k})=-\frac{3}{4k^2}\frac{\partial\delta_{\gamma\mathbf{k}}}{\partial\eta},
\label{eq:b7}
\nonumber
\end{eqnarray}
where according to (\ref{eq:16}), $\delta_\gamma=4(\Psi+\mathcal{E}/3)$.
By $\Phi_{\mathbf{k}}$ and the invariant radiation energy contrast
$\delta_{\gamma\mathbf{k}}$ we denote amplitudes
of corresponding perturbations as plane waves, which are functions only of the conformal time $\eta$.
We have omitted the integrated Sachs--Wolfe effect here, since its contribution
for high multipole moments is negligible and for low multipole moments it was
computed by {{\it Balek and Skovran}, [2015]}.

Relative anisotropies in the CMB temperature can be expanded into spherical harmonics,
with the coefficients that can be written as scalar product of the anisotropy with
spherical harmonics.
\begin{eqnarray}
\frac{\delta T}{T}(\eta_0,\mathbf{l})=\sum_{lm}{a_{lm}Y_{lm}(\mathbf{l})},
\qquad
a_{lm}=\int d^2\mathbf{l}\frac{\delta T}{T}(\eta_0,\mathbf{l})Y^{*}_{lm}(\mathbf{l}).
\label{eq:c1}
\end{eqnarray}
Since the CMB anisotropies are random, the coefficients $a_{lm}$ are random as well.
In order to compare the theory with the observed data, correlation functions must be introduced.
The two-point correlation function is defined as
$C(\vartheta)\equiv\left< T^{-2} \delta T (\eta_0,\mathbf{l}_1) \delta T (\eta_0,\mathbf{l}_2) \right>$,
where $\vartheta$ is the angle between $\mathbf{l}_1$ and $\mathbf{l}_2$ and the brackets
$\left<\textrm{ }\right>$
denote averaging over observer's position. This function can be written as a sum over multipole moments,
\begin{eqnarray}
C(\vartheta)=\frac{1}{4\pi}\sum_l{(2l+1) \left<|a_{lm}|^2\right> P_l(\cos\vartheta)},
\nonumber
\label{eq:c2}
\end{eqnarray}
where $C_l=\left<|a_{lm}|^2\right>$ are the coefficients of the angular power spectrum of CMB
and $P_l$ are Legendre polynomials.
The mean values $\left<|a_{lm}|^2\right>$ are independent on $m$ because there is no preferred direction in the universe,
but for the given observer there exists statistical randomness of the CMB anisotropies,
so that the observed $|a_{lm}|^2$
are not equal for all $m$ for the given multipole moment $l$.
The best estimate of the values of coefficients of angular power spectrum is then given by averaging
\begin{eqnarray}
C_l = \frac{1}{2l+1}\sum_{m=-l}^l|{a_{lm}|^2},
\label{eq:c3}
\nonumber
\end{eqnarray}
with an unavoidable error known as the {\it cosmic variance}
\begin{eqnarray}
\frac{\Delta C_l}{C_l} = \sqrt{\frac{2}{2l+1}}.
\nonumber
\label{eq:c4}
\end{eqnarray}
Using (\ref{eq:b10}) and (\ref{eq:c1}) we find
\begin{eqnarray}
& &C_l = \left< |a_{lm}|^2 \right> = \frac{2}{\pi} \int\limits_0^{\infty} \tau_l(k)^2 (2\pi)^3 \mathcal{P}_{\Phi}(k) k^2 dk,
\label{eq:c5}
\\
& &\tau_l(k)\equiv
\left[\frac{\alpha(\mathbf{k})}{\Phi^{(0)}_{\mathbf{k}}}j_l(k(\eta_0-\eta_r))
+ \frac{\beta(\mathbf{k})}{\Phi^{(0)}_{\mathbf{k}}}\frac{\partial}{\partial\eta_0}j_l(k(\eta_0-\eta_r))
\right] e^{-\Sigma k^2},
\nonumber
\end{eqnarray}
where $j_l$ are spherical Bessel functions of the first kind and
the function of the comoving wavenumber
$\mathcal{P}_{\Phi}(k)$ known as {\it power spectrum} is defined by
\begin{eqnarray}
\left<\Phi^{(0)}_{\mathbf{k}}\Phi^{(0)}_{\mathbf{k}'}\right>
= (2\pi)^3 \delta^{(3)}(\mathbf{k}+\mathbf{k}') \mathcal{P}_{\Phi}(k).
\nonumber
\label{eq:c6}
\end{eqnarray}
The function $\tau_l(k)$ is called transfer function.
It depends only on the comoving wavenumber $k$
and not on the direction of the vector $\mathbf{k}$ because
the equations governing the evolution of perturbations are isotropic.
The power spectrum is usually written as $\mathcal{P}_{\Phi}(k) \propto k^{n_s-4}$
where $n_s$ is the {\it spectral index} given by inflationary models.
The variance of the coefficients of angular power spectrum due to presence of
the radiation-like solid is defined as
\begin{eqnarray}
\frac{\Delta_\xi C_l}{C_l} \equiv \frac{C_l(\xi)-C_l}{C_l},
\nonumber
\label{eq:c16}
\end{eqnarray}
where $C_l$ and $C_l(\xi)$ are the coefficients of angular power spectrum
without the presence of the radiation-like solid and with its presence respectively.
The values of $\Delta_\xi C_l / C_l$ are compared with the cosmic variance
for the {\it Planck} values of cosmological parameters in Figure~1.
The graph was plotted with the use of integral (\ref{eq:c5}) where $\alpha(\mathbf{k})$
and $\beta(\mathbf{k})$ were calculated by solving differential equations
(\ref{eq:10})-(\ref{eq:13}) numericaly for $\xi=10^{-4}$.
Calculation of $\Delta_\xi C_l / C_l$ for different values of shear modulus parameter $\xi$
revealed that for high multipole moments $\Delta_\xi C_l / C_l$ can be well approximated as
a linear function of $\xi$.
The presence of the radiation-like solid can be confirmed by
observations only if $|\Delta_\xi C_l/C_l| \gtrsim |\Delta C_l/C_l|$,
which can be satisfied not only for very low multipole moments, as shown by {{\it Balek and Skovran} [2015]},
but also for very high ones.

\begin{figure}[htb]
\centering
\includegraphics[scale=0.12]{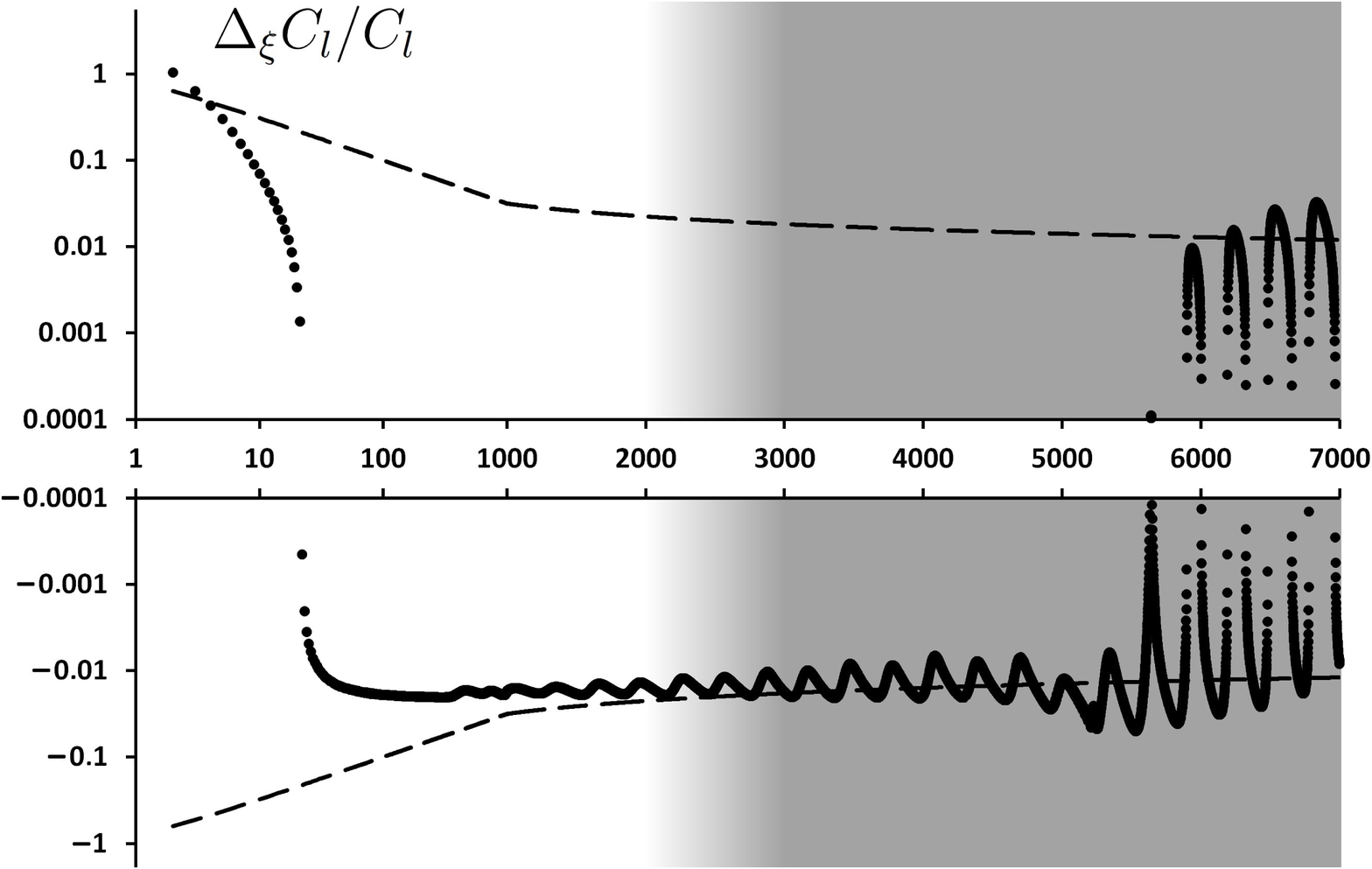}

\caption{The coefficients of angular power spectrum variance due to presence of
the radiation-like solid with $\xi = 10^{-4}$ (dots) exceed the cosmic variance
(dashed line) for very low multipole moments ($l<3$, if the integrated Sachs--Wolfe effect is not included)
and also for some very high
($l \gtrsim 3000$) ones (horizontal axis). We have used $k_{min}=0.01/\eta_{*}$
as the lower limit of integral (\ref{eq:c5}).
For the vertical axis we have used logarithmic scale for positive and negative values separately
and for the horizontal axis we have combined logarithmic scale (for $l$ up to $1000$)
with ordinary scale ($l$ over $1000$).
The grey area represents the sector which is beyond the reach of curent observations
\label{figure}}
\end{figure}

Due to the Silk damping, frequency and amplitude of oscillating functions of the wavenumber
$\alpha(\mathbf{k})/\Phi^{(0)}_\mathbf{k}$ and $\beta(\mathbf{k})/\Phi^{(0)}_\mathbf{k}$
depend on the wavenumber and, therefore, angular power spectrum variance $\Delta_\xi C_l / C_l$
behaves differently in different multipole moment sectors.
This is the reason why oscillations of $\Delta_\xi C_l / C_l$ shown in Figure~1 change
qualitatively for $l\approx5500$.

Note that $C_2(\xi)$ diverges due to the contribution of the long-wavelength part of
transfer-function $\tau_2(k)$ to the integral (\ref{eq:c5}). Because of finite duration of inflation,
the wavelength of the quantum fluctuations did not increase to infinity and the long-wavelength
part of transfer-function actually does not contribute to this integral.
Hence no divergence occures and integral (\ref{eq:c5}) must be computed with nonzero
lower limit $k_{min}$. The value of $k_{min}$ can be estimated as $k_{min}\eta_0\sim e^{N_{min}-N}$
where $N$ is the number of $e$-folds during inflation and $N_{min}$ is the minimal
value of $N$ needed to homogenize the universe at the scale of the Hubble radius.
This cutoff affects only the lowest coefficients of the angular power spectrum.

\section{Conclusion}
We have applied perturbation theory in the proper-time comoving gauge to a universe
filled with radiation, baryonic matter and CDM, and considered the presence of
a radiation-like solid with pressure to energy density ratio $w=1/3$ and
constant shear modulus to energy density ratio $\xi$.
The presence of such a solid does not change the evolution of the unperturbed universe.
In order to obtain results also for short-wavelength perturbations, we had to
extend the theory developed by {{\it Balek and Skovran} [2015]} and consider the CDM not coupled
with the baryonic matter, which leads to more complicated equations.

We have calculated effects of the radiation-like solid on the CMB angular power spectrum.
Its presence may have a significant effect
on it not only for very low multipole moments,
as shown by {{\it Balek and Skovran} [2015]}, but also for very high ones.
For $\xi\sim10^{-4}$, the effect can be observable for $l\gtrsim3000$,
while according to {{\it Balek and Skovran} [2015]}
the large-angle anisotropies are not affected significantly.
They considered also the integrated Sachs--Wolfe effect,
which is not included in this work,
because it affects the angular power spectrum significantly only for
the low multipole moments.
Our model is in agreement with current observations
for appropriate values of the shear modulus coefficient $\xi$
while causing effects not predicted by standard theory,
which are beyond the reach of current observations.

Since the number of affected coefficients of the CMB angular power spectrum
for very high multipole moments surpass their number for low multipole moments,
future observations of the CMB anisotropies having appropriate resolution
could confirm or refute the presence of the radiation-like solid with much greater certainty
than it is possible to do today.

\vskip 1.5cm \noindent
{\bf{\Large Acknowledgment}}
\vskip 5mm \noindent I would like to thank my supervisor Vladim\'ir Balek for helpful discussions and valuable advices.

\vskip 1.5cm \noindent
{\bf{\Large References}}

\vskip 5mm \noindent
Bucher N., Spergel D. N., Is the dark matter a solid? {\it Phys. Rev.} {\bf D60}, 043505 (1999)

\vskip 2mm \noindent
Battye R. A., Bucher N., Spergel D. N., Domain wall dominated universes, astro-ph/9908047 (1999)

\vskip 2mm \noindent
Leite A., Martins C., Scaling properties of domain wall networks, {\it Phys. Rev.} {\bf D84}, 103523 (2011)

\vskip 2mm \noindent
Battye R. A., Moss A., Anisotropic perturbations due to dark energy, {\it Phys. Rev.} {\bf D74}, 041301 (2006)

\vskip 2mm \noindent
Battye R. A., Moss A., Anisotropic dark energy and CMB anomalies, {\it Phys. Rev.} {\bf D80}, 023531 (2009)

\vskip 2mm \noindent
Battye R. A., Moss A., Cosmological Perturbations in Elastic Dark Energy Models, {\it Phys. Rev.} {\bf D76}, 023005 (2007)

\vskip 2mm \noindent
Battye R. A., Pearson J. A., Massive gravity, the elasticity of space-time and perturbations
in the dark sector, {\it Phys. Rev.} {\bf D88}, 084003 (2013)

\vskip 2mm \noindent
Kumar S., Nautiyal A., Sen A. A., Deviation from $\Lambda$CDM with cosmic strings networks, {\it Eur. Phys.} {\bf J. C73}, 2562 (2013)

\vskip 2mm \noindent
Gruzinov A., Elastic Inflation, {\it Phys. Rev.} {\bf D70}, 063518 (2004)

\vskip 2mm \noindent
Endlich S., Nicolis A., Wang J., Solid inflation, {\it JCAP10} (2013) 011

\vskip 2mm \noindent
Akhshik M., Clustering fossils in solid inflation, {\it JCAP} {\bf 1505}, 043 (2015).

\vskip 2mm \noindent
Bartolo N., Matarrese S., Peloso M. and Ricciardone A., Anisotropy in solid inflation, arXiv:1306.4160 [astro-ph.CO] (2013)

\vskip 2mm \noindent
Sitwell M., Sigurdson K., Quantization of Perturbations in an Inflating Elastic Solid, arXiv:1306.5762 [astro-ph.CO] (2013)

\vskip 2mm \noindent
Balek V., Skovran M., Cosmological perturbations in the presence of a solid with positive
pressure, arXiv:1401.7004 [gr-qc] (2014)

\vskip 2mm \noindent
Balek V., Skovran M., Effect of radiation-like solid on CMB anisotropies, {\it Class. Quant. Grav.} {\bf 32}, 015015 (2015)

\vskip 2mm \noindent
Pol\'ak V., Balek V., Plane waves in a relativistic homogeneous and isotropic elastic
continuum, {\it Class. Quant. Grav.} {\bf 25}, 045007 (2008)

\vskip 2mm \noindent
Mukhanov V., Physical Foundations of Cosmology, CUP, Cambridge (2005)

\end{document}